\documentclass[aps,prl,amsmath,twocolumn,showpacs]{revtex4-1}
\usepackage{amsmath, amssymb, graphicx, bm}
\usepackage[usenames,dvipsnames,svgnames,table]{xcolor}
\usepackage[colorlinks=true,linkcolor=RoyalBlue,citecolor=RoyalBlue]{hyperref}

\begin{document}
\title{Terahertz Antiferromagnetic Spin Hall Nano-oscillator}

\author{Ran Cheng}
\email{rancheng@utexas.edu}

\author{Di Xiao} 
\affiliation{Department of Physics, Carnegie Mellon University, Pittsburgh, PA 15213}

\author{Arne Brataas}
\affiliation{Department of Physics, Norwegian University of Science and Technology, NO-7491 Trondheim, Norway}


\begin{abstract}
We consider the current-induced dynamics of insulating antiferromagnets in a spin Hall geometry. Sufficiently large in-plane currents perpendicular to the N\'{e}el order trigger spontaneous oscillations at frequencies between the acoustic and the optical eigenmodes. The direction of the driving current determines the chirality of the excitation. When the current exceeds a threshold, the combined effect of spin pumping and current-induced torques introduces a dynamic feedback that sustains steady-state oscillations with amplitudes controllable via the applied current. The ac voltage output is calculated numerically as a function of the dc current input for different feedback strengths. Our findings open a route towards terahertz antiferromagnetic spin-torque oscillators. 
\end{abstract}

\maketitle

\textit{Introduction.}---The discovery of spin-transfer torques (STTs)~\cite{ref:STT1,ref:STT2} initiated an intense search for current-induced phenomena in magnetic materials because a STT can compensate the magnetic damping and induce spontaneous magnetization dynamics. When such a compensation occurs, the magnetization either switches to another direction~\cite{ref:SHswitch_Ralph1,ref:SHswitch_Ralph2} or evolves into a steady-state oscillation~\cite{ref:Tsoi,ref:Chien,ref:STNO_Ralph,ref:STNO_Silva,ref:STNO_Zeng}. While the former improves writing operations in magnetic memory devices, the latter enables sustainable ac signal generation from dc inputs, known as spin-torque oscillators~\cite{ref:STOreview,ref:chapter38}. In ferromagnets, currents or magnetic fields can tune the output frequency in the range from the megahertz to the gigahertz regime.

Spin-torque oscillators can potentially be operated at much higher terahertz frequencies when antiferromagnets (AFs) replace ferromagnets. Two facts make this possible: (1) the eigenfrequencies of typical AFs fall into the terahertz range~\cite{ref:AFMR} and (2) a STT can trigger spontaneous excitations of an AF in a similar way as ferromagnets~\cite{ref:SPAF,ref:Matthew,ref:Gomonay}. While most AFs are insulators preventing the STTs to be operative by passing a current through the sample, the spin Hall effect (SHE)~\cite{ref:SHE} is an alternative that generates STTs even when electrons do not flow through the magnet~\cite{ref:SHNO_Saitoh,ref:SHNO_Ralph}. The latter phenomenon provides an avenue towards low-dissipation spin Hall nano-oscillators (SHNOs)~\cite{ref:SHNO_Urazhdin,ref:SHNO_simulation}.

However, to realize AF-based SHNOs, current-induced excitations should not grow indefinitely. Instead, they should evolve into steady-state oscillations with substantial output power~\cite{note}. Although an AF does not suffer magnetic switching even when a STT overcomes the damping, its N\'{e}el vector will directly evolve into a right-angle precession around the direction of the spin accumulation~\cite{ref:Gomonay}. Since the amplitude of such a dynamical motion is not continuously tunable via the applied current, it does not meet the requirements of a SHNO. 

Steady-state oscillations are realizable in ferromagnets for the following reasons. According to the original form of the STT~\cite{ref:STT1}, its angle dependence and that of the Gilbert damping differ~\cite{ref:JXangle}. Therefore, as the amplitude of a spontaneous motion is growing, the two competing mechanisms will balance each other at a unique angle---that is where a steady-state oscillation takes place. Nevertheless, this feature is not active when the SHE operates the STT. In the latter scenario, one needs to introduce alternative mechanisms to prevent a spontaneous excitation from growing into magnetic switching. For example, the spatially localized mode~\cite{ref:Bullet} produces a nonlinearity that can sustain its auto-oscillations~\cite{ref:SHNO_Urazhdin}. To remain within spatially uniform excitations, the dipolar interaction is often required~\cite{ref:Rezende}. However, the dipolar interaction is negligible in AFs where the magnetization is vanishingly small. 

In this Letter, we exploit a recently proposed feedback mechanism~\cite{ref:Feedback} to realize a terahertz SHNO in an AF/heavy-metal heterostructure. The feedback originates from the combined effect of the SHE and its reverse process, which connects spin pumping with the spin backflow~\cite{ref:spinbattery,ref:backflow}. It is entirely independent of the dipolar interaction. First, we determine the threshold of spontaneous excitations by solving the N\'{e}el order dynamics in the linear response regime and relate the threshold to a current density. Then, we numerically explore the nonlinear N\'{e}el order dynamics beyond the threshold by including the feedback effect~\cite{ref:Feedback} and show that the feedback is indispensable to maintain uniform auto-oscillations. Finally, we demonstrate that in contrast to previous studies~\cite{ref:Gomonay}, our proposed SHNO creates a substantial ac voltage output with its amplitude continuously tunable via the applied dc current.

\textit{Dynamics.}---We assume that the AF has a single crystal structure, and describe the sublattice magnetizations by two unit vectors $\bm{m}_A$ and $\bm{m}_B$. We introduce the N\'{e}el vector $\bm{\ell}=(\bm{m}_A-\bm{m}_B)/2$, and the small magnetization $\bm{m}=(\bm{m}_A+\bm{m}_B)/2$; they satisfy $\bm{m}\cdot\bm{\ell}=0$ and $m^2+\ell^2=1$. In the exchange limit, $m\ll1$, thus $\ell^2\approx1$ and $\bm{\ell}\cdot\dot{\bm{\ell}}=0$. The Cartesian coordinates are chosen such that the hard axis is along $\bm{\hat{z}}$, and the in-plane easy-axis along $\bm{\hat{x}}$. We scale everything in (positive) angular frequency, where the hard axis anisotropy is described by $\omega_{\perp}$, the easy in-plane anisotropy $\omega_{\parallel}$, and the Heisenberg exchange interaction $\omega_E$. In the macrospin description, the free energy is $F=-\hbar\omega_E\bm{\ell}^2-\hbar\omega_{\parallel}[(\bm{\hat{x}}\cdot\bm{\ell})^2+(\bm{\hat{x}}\cdot\bm{m})^2]/2+\hbar\omega_{\perp}[(\bm{\hat{z}}\cdot\bm{\ell})^2+(\bm{\hat{z}}\cdot\bm{m})^2]/2$, which defines two thermodynamic forces $\hbar\bm{f}_{\ell}=-\partial F/\partial\bm{\ell}$ and $\hbar\bm{f}_m=-\partial F/\partial\bm{m}$~\cite{ref:phenom}. The coupled equations of motion are
\begin{subequations}
	\label{eq:dynamics}
	\begin{align}
		\dot{\bm{m}} &=\bm{f}_{\ell}\!\times\!\bm{\ell}+\bm{f}_m\!\times\!\bm{m}+\alpha(\bm{m}\!\times\!\dot{\bm{m}}+\bm{\ell}\!\times\!\dot{\bm{\ell}})+\bm{\tau}_{m}\ , \label{eq:mLLG}\\
		\dot{\bm{\ell}}&=\bm{f}_{m}\!\times\!\bm{\ell}+\bm{f}_{\ell}\!\times\!\bm{m}+\alpha(\bm{m}\!\times\!\dot{\bm{\ell}}+\bm{\ell}\!\times\!\dot{\bm{m}})+\bm{\tau}_{\ell}\ , \label{eq:nLLG}
	\end{align}
\end{subequations}
where $\alpha$ is the Gilbert damping constant, and $\bm\tau_\ell$ and $\bm\tau_m$ are the STTs given by Ref.~\cite{ref:SPAF,ref:Matthew,ref:Gomonay}
\begin{subequations}
	\label{eq:STTs}
	\begin{align}
		&\bm{\tau}_{m}=\bm{\ell}\times(\bm{\omega}_s\times\bm{\ell})+\bm{m}\times(\bm{\omega}_s\times\bm{m})\ , \label{eq:taum}\\
		&\bm{\tau}_{\ell}=\bm{\ell}\times(\bm{\omega}_s\times\bm{m})+\bm{m}\times(\bm{\omega}_s\times\bm{\ell})\ . \label{eq:taun}
	\end{align}
\end{subequations}
Here, $\bm{\omega}_s$ is the vector of spin accumulation; its magnitude (in frequency units) represents the STT strength. 

\begin{figure}[t]
	\centering
	\includegraphics[width=\linewidth]{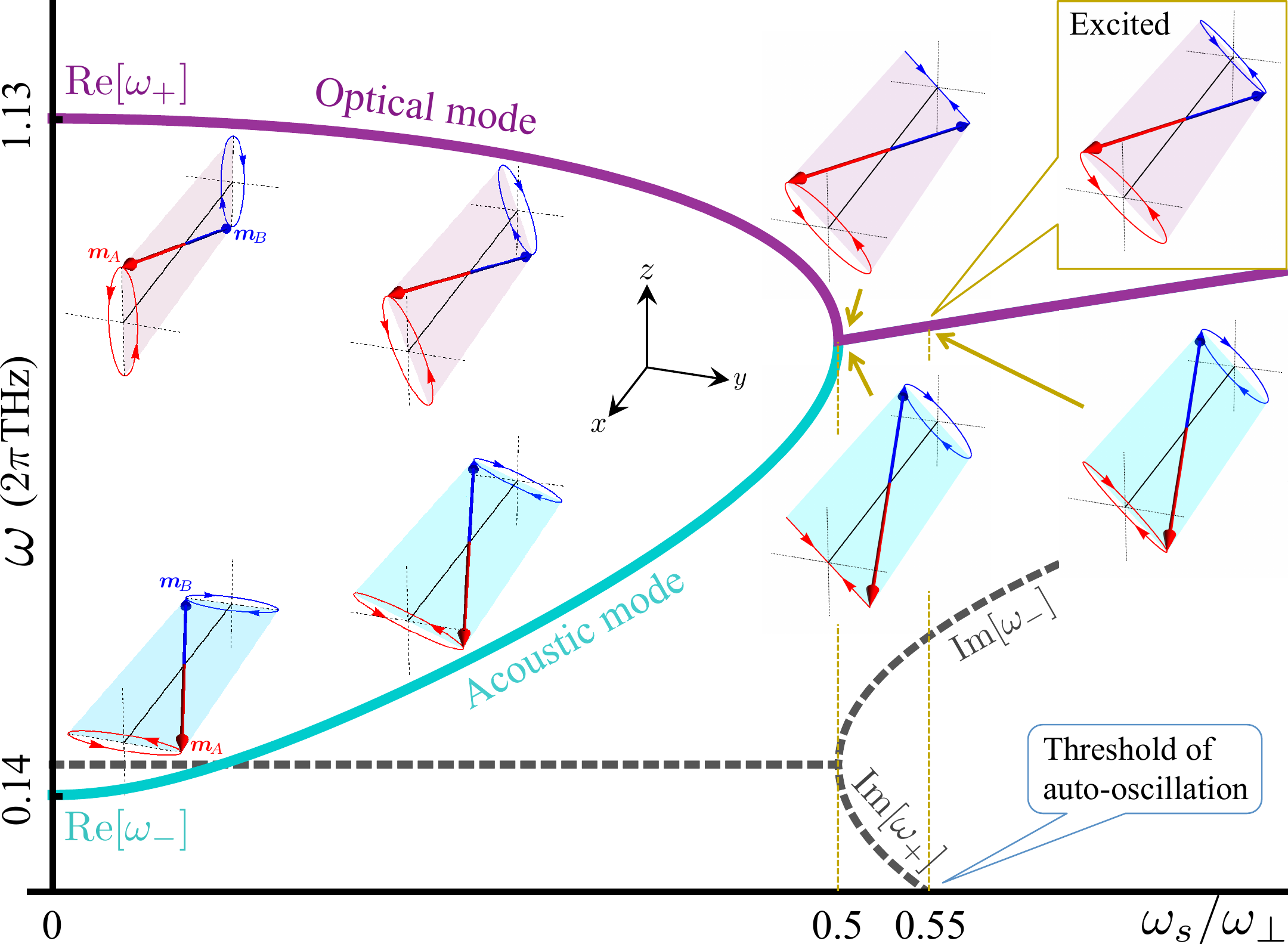}
	\caption{(Color online) Evolutions of the eigenfrequencies and the eigenmodes of NiO~\cite{ref:NiO} with an increasing STT. In the region $\omega_s<\omega_{\perp}/2$, $\mathrm{Re}[\omega_+]$ and $\mathrm{Re}[\omega_-]$ approach each other with the increasing STT while $\mathrm{Im}[\omega_{\pm}]$ remains degenerate. Both $\bm{m}_A$ and $\bm{m}_B$ precess elliptically with opposite chiralities. As $\omega_s$ increases, the long axis of the optical (acoustic) precession tilts away from the z-axis (y-axis). At $\omega_s=\omega_{\perp}/2$, $\mathrm{Re}[\omega_{+}]$  and $\mathrm{Re}[\omega_{-}]$ become degenerate while $\mathrm{Im}[\omega_{+}]$ and $\mathrm{Im}[\omega_{-}]$ separate; the elliptical orbit traversed by $\bm{m}_B$ ($\bm{m}_A$) in the optical (acoustic) mode shrinks into a line and then opens up into an ellipse again, with its chirality changing sign. Hence for $\omega_s>\omega_{\perp}/2$, $\bm{m}_A$, $\bm{m}_B$, and $\bm{\ell}$ in the optical (acoustic) mode all have the right-handed (left-handed) chirality. At $\omega=0.55\omega_{\perp}$, $\mathrm{Im}[\omega_+]$ vanishes, the optical mode is excited, and the linear response breaks down. }
	\label{fig:mode}
\end{figure}

To derive the current-induced excitations, we decompose the N\'{e}el vector as $\bm{\ell}=\hat{\bm{x}}+\bm{\ell}_{\perp}e^{i\omega t}$, assuming $|\bm{\ell}_{\perp}|\ll1$. Restricting to linear order in $\bm{\ell}_{\perp}$, we can eliminate $\bm{m}$ in Eqs.~\eqref{eq:mLLG} and~\eqref{eq:nLLG}, and obtain the eigenfrequencies as $\omega_{\pm}/\omega_E=i\alpha+\left[(\omega_{\perp}+2\omega_{\parallel}\pm\sqrt{\omega_{\perp}^2-4\omega_s^2})/\omega_E-\alpha^2\right]^{1/2}$, where the $+$ ($-$) sign corresponds to the optical (acoustic) mode~\cite{supp}. In Fig.~\ref{fig:mode}, we plot the two eigenfrequencies as a function of the STT strength $\omega_s$ with parameters taken from NiO~\cite{ref:NiO}. As $\omega_s$ increases, the real parts $\mathrm{Re}[\omega_{+}]$ and $\mathrm{Re}[\omega_{-}]$ approach each other until they become degenerate at $\omega_s=\omega_{\perp}/2$. By contrast, the imaginary parts $\mathrm{Im}[\omega_{+}]$ and $\mathrm{Im}[\omega_{-}]$ remain degenerate and unaffected for $\omega_s<\omega_{\perp}/2$. But when $\omega_s>\omega_{\perp}/2$, $\mathrm{Im}[\omega_+]$ ($\mathrm{Im}[\omega_-]$) reduces (grows) rapidly, indicating that the damping is diminished (enhanced) by the STT. At the threshold~\cite{ref:Gomonay,supp}
\begin{align}
 \omega_s^{\mathrm{th}}=\sqrt{\frac{\omega_{\perp}^2}{4}+\alpha^2(2\omega_{\parallel}+\omega_{\perp})\omega_E}\ , \label{eq:thresholdomega}
\end{align}
$\mathrm{Im}[\omega_+]$ vanishes, which marks the onset of spontaneous excitation of the optical mode and the breakdown of the linear response approximation. The uniaxial symmetry enforces that $\mathrm{Im}[\omega_+]$ also vanishes for $-\omega_s^{\mathrm{th}}$ so that the auto-oscillation can be triggered by a reversed current as well. Hereafter, we restrict our discussions to positive $\omega_s$ unless otherwise stated.

In the absence of the hard-axis anisotropy ($\omega_{\perp}=0$), the threshold~\eqref{eq:thresholdomega} is linear in $\alpha$, so the antidamping effect occurs when the STT is turned on. However, in the general case where $\omega_{\perp}>0$, the antidamping effect appears only when $\omega_s>\omega_{\perp}/2$ as shown by the $\mathrm{Im}[\omega_+]$ curve in Fig.~\ref{fig:mode}, whereas a driving STT in the regime $\omega_s<\omega_{\perp}/2$ modifies the patterns of the eigenmodes as illustrated in Fig.~\ref{fig:mode} (also see Ref.~\cite{ref:Gomonay}). Specifically, an increasing STT drags the long axes of the elliptical precessions away from their original orientations until they are 45$^{\circ}$ away from the hard-axis. In spite of this change, $\bm{m}_A$ and $\bm{m}_B$ always exhibit opposite chiralities, \textit{i.e.}, as seen from the $x$-direction, $\bm{m}_A$ ($\bm{m}_B$) rotates counterclockwise (clockwise). However, at the degenerate point $\omega_s=\omega_{\perp}/2$, the chirality of $\bm{m}_B$ ($\bm{m}_A$) in the optical (acoustic) mode reverses. Consequently, when $\omega_s>\omega_{\perp}/2$, both $\bm{m}_A$ and $\bm{m}_B$, hence the N\'{e}el vector $\bm{\ell}$, all acquire the same chirality. At the threshold $\omega_s^{\mathrm{th}}$, the excited optical mode is right-handed. If $\omega_s$ changes sign, the excited mode is still the optical mode, but its chirality becomes left-handed. These observations suggest that the direction of the current determines the chirality of the excitation.

\begin{figure}[t]
	\centering
	\includegraphics[width=0.92\columnwidth]{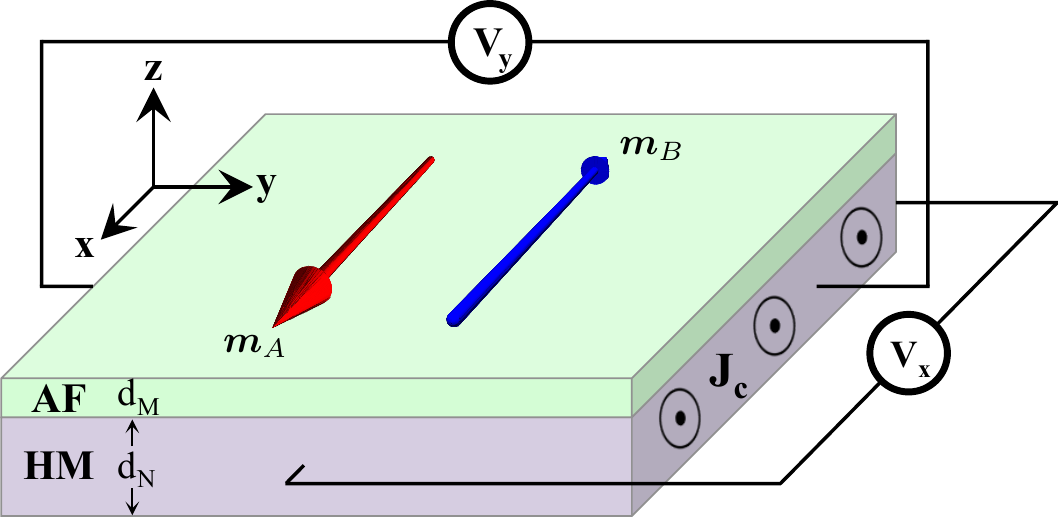}
	\caption{(Color online) An insulating AF/HM heterostructure. The applied dc current density $\bm{J}_c$ drives the AF via the SHE. The dynamics of the AF pumps spin current back into N, and converts into electric field via the inverse SHE, which is monitored by two voltmeters.}
	\label{fig:device}
\end{figure}

\textit{Critical current.}---Consider a setup consisting of an insulating AF deposited on a heavy-element normal metal (HM) with spin-orbit coupling, as shown schematically in Fig.~\ref{fig:device}. We assume a current density $\bm{J}_c$ is applied along the $y$-direction; it is perpendicular to the N\'{e}el vector of the AF. The SHE in the HM generates antidamping STTs to drive the N\'{e}el vector dynamics, which in turn pumps spin current back into the HM. The pumped spin current converts into a charge voltage due to the inverse SHE~\cite{ref:ISHE}, which is detected by two voltmeters. Let $d_M$ be the thickness of the AF, $d_N$ the thickness of the HM, and assume that the HM has spin diffusion length $\lambda$, lattice constant $a$, and conductivity $\sigma$. By solving the spin diffusion equation in the presence of the SHE~\cite{ref:Feedback,ref:JX} under boundary conditions involving both spin pumping and STTs~\cite{ref:SPAF,supp}, we relate the threshold STT Eq.~\eqref{eq:thresholdomega} to a critical current density
\begin{align}
 J_c^{\mathrm{th}}=\omega_s^{\mathrm{th}}\frac{d_M(h\sigma+2\lambda e^2g_r\coth\frac{d_N}{\lambda})}{2\theta_s a^3\lambda e g_r\tanh\frac{d_N}{2\lambda}}\ , \label{eq:thresholdsolution}
\end{align}
where $\theta_s$ is the spin Hall angle, $-e$ is the electron charge, and $g_r$ is the areal density of the transverse mixing conductance~\cite{ref:SPAF}. From Eq.~\eqref{eq:thresholdsolution}, we see that the critical current density $J_c^{\mathrm{th}}$ can be lowered by reducing (increasing) the thickness of the AF $d_M$ (HM $d_N$). For example, consider a NiO(1)/Pt(25) (numbers in nm) bilayer. At room temperature, we use Pt material parameters from Ref.~\cite{ref:Ptdata}, and use $g_r \approx 1.2e^2/h$ per $a^2$ for perfect interfaces~\cite{ref:SPAF}. Since $a=0.417$nm in NiO, Eq.~\eqref{eq:thresholdsolution} gives $J_c^{\mathrm{th}}=2.91\times10^8$A/cm$^2$. For uniaxial AFs such as MnF$_2$, the hard-axis anisotropy is absent ($\omega_{\perp}=0$), thus $\omega_s^{\mathrm{th}}$ will be appreciably smaller, so will $J_c^{\mathrm{th}}$.

In real AF/HM heterostructures, the critical current density could be higher than the above estimation since the surface roughness can diminish the transverse mixing conductance $g_r$. Nevertheless, a large spin Hall angle by using \textit{e.g.} topological insulators~\cite{ref:TI} can reduce the critical current. In addition, even though domain formation can happen, the N\'{e}el vector $\bm{\ell}$ survives a spatial average over all domains since it is bi-axial (for comparison, the magnetization vector of a ferromagnet is uniaxial). In this sense, the domain formation amounts to a reduction of the volume density of the N\'{e}el vector. While these imperfections renormalize the material parameters, they do not qualitatively invalidate the essential physics.

\begin{figure}[t]
	\centering
	\includegraphics[width=0.82\columnwidth]{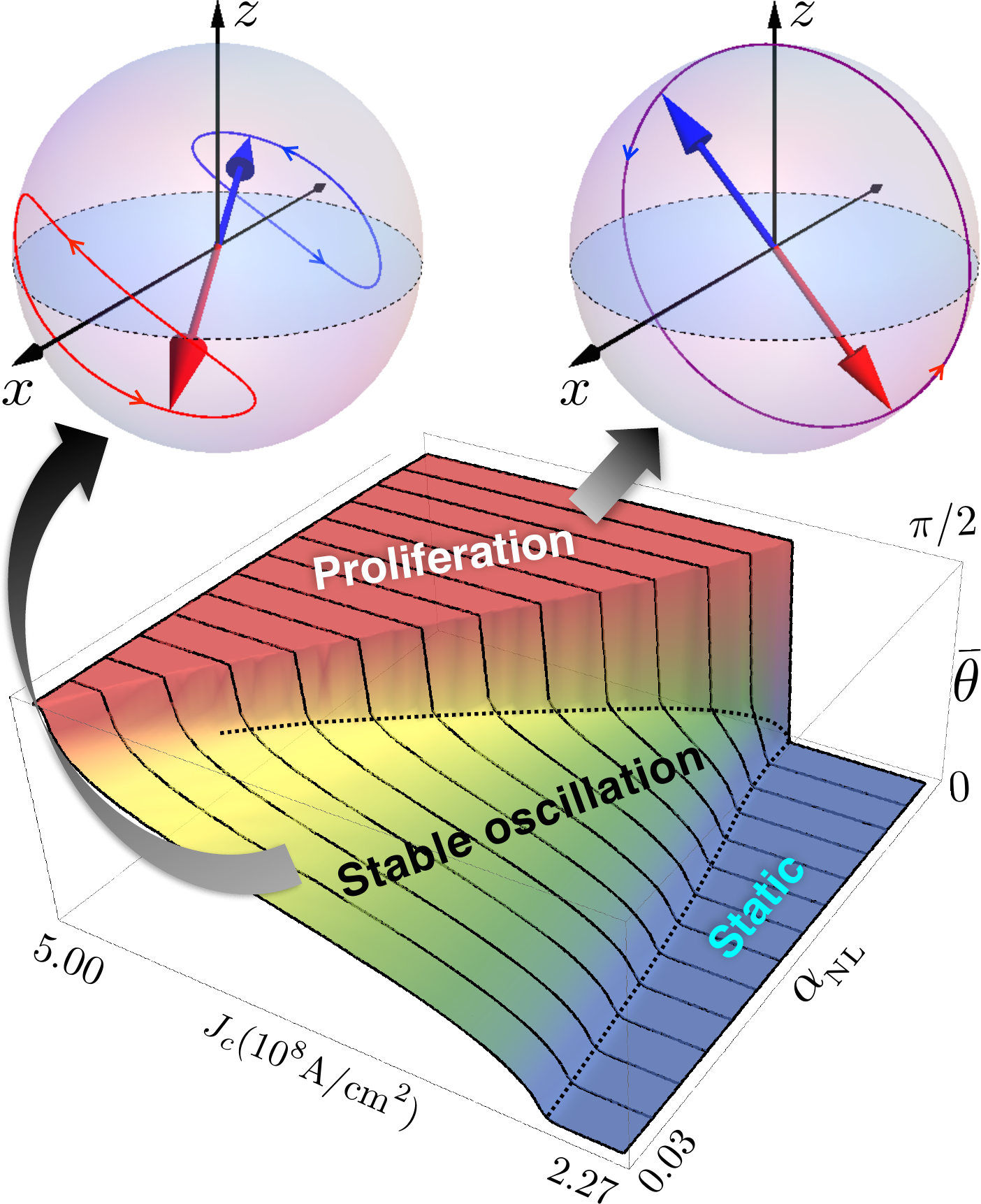}
	\caption{(Color online) Phase diagram of a SHNO based on the NiO(1)/Pt(25) bilayer structure. Phases are characterized by the numerical result of the time-averaged terminal angle $\bar{\theta}\equiv\frac1T\lim\limits_{t\rightarrow\infty}\int_t^{t+T}dt'\arcsin|\ell_{\perp}(t')|$ as a function of the applied current density $J_c(10^8$A/cm$^2)\in[2.27,5.00]$ and the feedback strength $\alpha_{_{\mathrm{NL}}}\in[0,0.03]$. Upper panels: the stable oscillation phase (left) differs from the proliferation phase (right) in that the orbits of $\bm{m}_A$ and $\bm{m}_B$ do not overlap, and the cone angle of the N\'{e}el vector $\theta<\pi/2$.}
	\label{fig:phase}
\end{figure}

\textit{Feedback.}---The linear response only allows us to solve the eigenmodes as those depicted in Fig.~\ref{fig:mode} and to predict the threshold of auto-oscillation excitations. Beyond the threshold, however, the assumption $|\bm{\ell}_{\perp}|\ll1$ is invalid, and we need to consider nonlinear responses. But in our calculations so far, both the Gilbert damping and the antidamping STT are linear in $\bm{\ell}_{\perp}$, so is the total effective damping. This behavior implies that the amplitude of a uniform excitation will grow exponentially with time since $\mathrm{Im}[\omega_+]<0$. In ferromagnets, this means that the magnetization switches to the opposite direction without any steady-state motion at intermediate configurations. Here in a collinear AF, the terminal status of the N\'{e}el vector is a right-angle precession around $\hat{\bm{x}}$ as shown in the upper-right panel of Fig.~\ref{fig:phase}, whereas oscillations at cone angles $\theta\in(0,\pi/2)$ are unstable.

\begin{figure*}[t]
	\centering
	\includegraphics[width=0.97\linewidth]{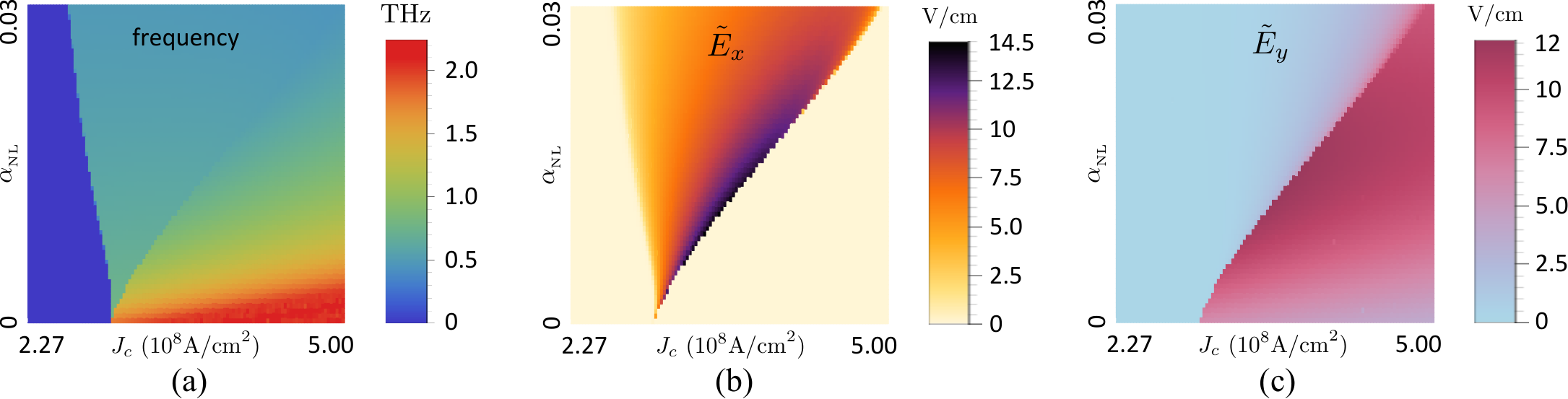}
	\caption{(Color online) Numerical results of multiple ac outputs associated with steady-state oscillations (at $t\rightarrow\infty$) in the range $J_c(10^8$A/cm$^2)\in[2.27,5.00]$ and $\alpha_{_{\mathrm{NL}}}\in[0,0.03]$ for a NiO(1)/Pt(25) heterostructure. (a): frequency output. (b) and (c): two perpendicular ac components (in effective values) of the electric field generated by the SHE and the spin pumping.}
	\label{fig:output}
\end{figure*}

However, the above analysis is incomplete as it ignores a crucial feedback effect~\cite{ref:Feedback}. The pumped spin current from a precessing N\'{e}el vector into the HM experiences a backflow that introduces a spin battery effect~\cite{ref:spinbattery}. In HMs, however, the spin pumping and the spin backflow are also connected via the combined effect of the SHE and its inverse process, which feeds the N\'{e}el vector dynamics back into itself. In ferromagnets, such a feedback mechanism manifests as a \textit{nonlinear} damping effect in the magnetization dynamics~\cite{ref:Feedback}. Following the same spirit, we can derive a similar feedback-induced damping effect for AFs in the exchange limit where $|\bm{m}|\ll|\bm{\ell}|$. To this end, recall that the pumped spin current into the HM converts into an electric field $\bm{E}$ due to the inverse SHE. According to Ohm's law, $\bm{J}_c=\sigma\bm{E}-\theta_s(\sigma/2e)\hat{\bm{z}}\times\partial_z\bm{\mu}_s$ where $\bm{\mu}_s$ is the spin accumulation in the HM. As we fixed the current density $\bm{J}_c$ through external circuits, a change of the electric field $\bm{E}$ necessarily leads to a change of the spin accumulation $\bm{\mu}_s$. Subsequently, the change of $\bm{\mu}_s$ diffuses and generates an additional spin current, which will finally deliver the influence of spin pumping back into the N\'{e}el vector through STTs. By closing such a feedback loop~\cite{supp}, we obtain a feedback torque that should be added to Eq.~\eqref{eq:taum} as
\begin{align}
	\bm{\tau}_{_{\mathrm{FB}}}=\alpha_{_{\mathrm{NL}}}[\ell_z^2\bm{\ell}\times\dot{\bm{\ell}}-\dot{\ell}_z(\hat{\bm{z}}\times\bm{\ell})]\ , \label{eq:FB}
\end{align}
where the feedback coefficient is
\begin{align}
	\alpha_{_{\mathrm{NL}}}=\theta_s^2\frac{a^3}{d_M}\frac{2\hbar\sigma\lambda e^2g_r^2\coth\frac{d_N}{\lambda}}{(h\sigma+2\lambda e^2g_r\coth\frac{d_N}{\lambda})^2}\ .
\end{align}
For the NiO(1)/Pt(25) bilayer structure considered earlier, $\alpha_{_{\mathrm{NL}}}=1.8\times10^{-4}$. When manipulating the material parameters, $\alpha_{_{\mathrm{NL}}}$ has a maximum $\theta_s^2g_ra^3/(8\pi d_M)$ at $h\sigma=2\lambda e^2g_r\coth d_N/\lambda$. While the feedback effect seems to be a higher order effect as $\alpha_{_{\mathrm{NL}}}$ is proportional to $\theta_s^2$, it can be significantly enhanced by searching for materials with large $\theta_s$. For example, it was shown recently that a topological insulator can potentially exhibit an extraordinarily large $\theta_s$ even greater than unity~\cite{ref:TI}. 

The feedback-induced nonlinear damping is a critical ingredient because it dramatically modifies the dynamical behavior of a SHNO using AFs. We demonstrate its effect by performing a numerical simulation with the result shown in Fig.~\ref{fig:phase}. For a given set of $(J_c,\ \alpha_{_{\mathrm{NL}}})$, we first run the simulation for a sufficiently long time so that the oscillation no longer grows. Then, we take a time average of the cone angle $\theta=\arcsin|\ell_{\perp}|$ over several periods. If we ignore the feedback, $\alpha_{_{\mathrm{NL}}}=0$, the AF either experiences no oscillation or proliferates into the right-angle precession ($\theta=\pi/2$). In the latter case, the orbits of $\bm{m}_A$ and $\bm{m}_B$ are circles and overlap each other completely (but their phases are still different by $\pi$). In the presence of the feedback, $\alpha_{_{\mathrm{NL}}}>0$, a finite window of stable oscillations at $\theta\in(0,\pi/2)$ opens up; the larger the $\alpha_{_{\mathrm{NL}}}$, the wider the window. In this novel phase, the terminal angle $\theta(t\rightarrow\infty)$ increases with increasing STT strength $\omega_s$ (hence $\bm{J}_c$). But at a sufficiently large $\omega_s$, the oscillator inevitably jumps into the proliferation phase, which marks a phase boundary separating stable oscillations from the proliferated configuration. However, this phase boundary terminates at extremely large $\alpha_{_{\mathrm{NL}}}$, after which transitions between the stabilized and proliferated oscillations are continuous.

\textit{Output.}---A salient feature of the novel stable oscillation phase is that the applied dc current density $\bm{J}_c$ controls the output power and that the output power is substantial: both features are indispensable for a SHNO. To quantify this fact, we explore multiple ac outputs as functions of $\bm{J}_c$ for different feedback strengths. Considering again the NiO(1)/Pt(25) structure, we run simulations for $J_c(10^8$A/cm$^2)\in[2.27,5.00]$ and $\alpha_{_{\mathrm{NL}}}\in[0,0.03]$. First, we plot the frequency output in Fig.~\ref{fig:output}(a). In the stable oscillation phase, the actual frequency output lies between the acoustic and the optical modes. Second, we study the ac voltage output from the inverse SHE and the spin pumping. For a fixed $\bm{J}_c$, the total electric field $\bm{E}=\bm{J}_c/\sigma+\Delta\bm{E}$ includes a time varying part~\cite{supp}
\begin{align}
 \Delta\bm{E}=\theta_s\frac{\hbar}{d_N}\frac{\lambda e g_r\tanh\frac{d_N}{2\lambda}}{h\sigma+2\lambda e^2 g_r\coth\frac{d_N}{\lambda}}(\bm{\ell}\times\dot{\bm{\ell}})\times\hat{\bm{z}}
\end{align}
that reflects our desired contribution. By eliminating the dc component of $\Delta\bm{E}$, we compute its ac components in the effective value $\widetilde{\bm{E}}=\lim\limits_{t\rightarrow\infty}\frac1T\int_t^{t+T}dt'|\Delta\bm{E}(t')|$ numerically. As shown in Figs.~\ref{fig:output}(b) and~(c), the $\widetilde{E}_x$ ($\widetilde{E}_{y}$) component is appreciably large (essentially zero) in the stable oscillation phase, whereas it vanishes (becomes nonzero) in the proliferation phase.  This contrasting property also enables a practical way to observe the phase transition. As illustrated in Fig.~\ref{fig:device}, the $\widetilde{\bm{E}}$ field is measured by two voltmeters. Since the maxima of both $\widetilde{E}_x$ and $\widetilde{E}_y$ reach 10 V/cm, the actual measured voltages from a nanometer-sized sample can be as large as $\mu$V.

\begin{acknowledgments}
 We are grateful to H. V. Gomonay, J. Xiao, M. W. Daniels, and J.-G. Zhu for insightful discussions. Work at Carnegie Mellon University is supported by the Department of Energy, Basic Energy Sciences, Grant No.~DE-SC0012509.
\end{acknowledgments}


\begin{thebibliography}{20}
\bibitem{note} The physics discussed here refers to the eigenmode excitation where the spin accumulation is parallel to the N\'{e}el vector, which should not be confused with the switching mode (Cheng \textit{et al.}, Phys. Rev. B \textbf{91}, 064423) where the spin accumulation is perpendicular to the N\'{e}el vector.
\end{thebibliography}
\end{document}